\documentclass[twocolumn,showpacs,preprintnumbers,amsmath,amssymb]{revtex4}

\usepackage{epsfig}
\usepackage{graphicx}
\usepackage{dcolumn}
\usepackage{bm}

\begin{document}

\title{Inverse Power Law Quintessence with Non-Tracking Initial Conditions}

\author{James P. Kneller}
\email{kneller@mps.ohio-state.edu}
\affiliation{Department of Physics, The Ohio State University, Columbus, Ohio 43210}
\affiliation{Department of Physics, North Carolina State University, Raleigh, North Carolina 27695-8202~\footnote{current address}}

\author{Louis E. Strigari}
\email{strigari@mps.ohio-state.edu}
\affiliation{Department of Physics, The Ohio State University, Columbus, Ohio 43210}

\date{\today}

\begin{abstract}

A common property of popular models of quintessence dark energy is the
convergence to a common solution from a large range of the initial conditions.
We re-examine the popular inverse power-law model of quintessence (where the common solution is
dubbed as the 'tracker') with particular attention to the initial conditions for the field and
their influence on the evolution.
We find that previously derived limits on the parameters of the potential in this model are
valid only in a range of initial conditions. A reasonably sharp boundary lies where the initial
energy density of the scalar field is equal to that of the background
radiation component. An initial quintessence energy density above this equipartition
value lead to a solution that will not have joined the tracker solution by the
present epoch. These non-tracker solutions possess the property that their present equation of
state is very compatible with the observed bounds and independent of the exponent of the potential.

\end{abstract}

\pacs{98.80.Cq}

\keywords{Dark Energy,Quintessence}

\maketitle


\section{Introduction}

Modern observations favor a dark component of the cosmological energy density
responsible for the present accelerating expansion rate of the Universe
\cite{SAG99}.  Quintessence is a proposed model for this component in which a
scalar field slowly evolves towards the minimum of its potential
\cite{RP88,W88,CDS98}. As the quintessence field, $Q$, nears this minimum
and the density of matter, $\rho_{M}$, simultaneously decreases, the expansion
rate will naturally start to accelerate. Making $Q$ dynamic alleviates the large
discrepancy between the energy densities of radiation, matter and dark
energy, most evident during the early Universe. The evolution of such a scalar
field depends entirely on its potential $V(Q)$. Several potentials have been
proposed and investigated, two in particular have gained interest because of their
simplicity: the inverse-power law (IPL), $V(Q)=V_{0}Q^{-\gamma}$
\cite{SWZ99,ZWS99,BR01,MG02,FW98}, and the exponential, $V(Q)=e^{-\lambda Q}$,
\cite{FJ98,CLW98}. Both of these potentials possess attractor-like solutions such
that a broad range of initial conditions converge to a unique solution at late
times \cite{RP88,LS99,J01}. Typically, during its evolution the ratio of the energy
density in $Q$ to the critical density of the universe, $\Omega_{Q}$, at a fixed
redshift is larger than the energy density of the pure cosmological constant it
hopes to replace, regardless of the potential used. In some cases, for example the
exponential potential, the energy density during the early Universe may be
large to affect Big Bang Nucleosynthesis (BBN)
\cite{CLW98,FJ98,KS03}. But in other models, such as the IPL, $\rho_{Q}$ is a
small and slowly increasing contributor of the energy density until close to
the present time when $\rho_{Q} \sim \rho_{M}$.

Probes like BBN, the Cosmic-Microwave Background (CMB), Type Ia Supernovae (SN Ia),
and Large Scale Structure (LSS) place restrictions on the parameters appearing in a
potential. By combining CMB, SN Ia, and LSS data, Bean \& Melchiorri obtained
a limit on the current equation of state of the field, $w_{Q}$, of
$-0.85 \geq w_{Q} \geq -1$ (at $1\sigma$) \cite{BM01}. In the IPL model
these results led to the conclusion that the exponent for the power
law, $\gamma$, must be $\alt 2$ at 95\% confidence \cite{ML02}. Similar
results were found in Corasaniti and Copeland \cite{CC02}, leading these authors to all but
rule out IPL models. Doran, Lilley, and Wetterich \cite{D02} combined CMB and LSS data and found that
IPL models with $\gamma \le 1$ were favored while Yahiro \emph{et al} \cite{YMIKO02} determined that BBN
can constrain tracker models with a conservative bound on the equation of state, $w_{Q} \le -0.2$, and a more stringent
bound on the initial energy density, $\rho_{Q}/\rho_{B} \lesssim 0.01$.
Yahiro \emph{et al} note that BBN provides a better constraint for models
in which $Q$ does not track at this epoch, limiting larger $\rho_{Q}$ in these models.

All these previous limits are derived from the assumption that $Q$ has reached its
tracker solution before the cosmological test is applied. In this paper we examine
the range of initial conditions of the IPL potential that lead to an evolution for
$Q$ that \emph{does not} converge to the tracker solution by the present epoch.
We begin by discussing the IPL model for
Quintessence and its tracker solution in section \ref{sec:tracker}, with
emphasis on the initial conditions of this solution. In section
\S\ref{sec:nontracker} we proceed to examine non-tracker solutions and
the behavior of $Q$ when the initial conditions vary significantly from the
tracker solution. Finally, in section \ref{sec:CMBSN Ia} we re-derive the
constraints on the parameters in the potential by using the CMB and SN Ia
observations when the tracker initial conditions are used, then we will examine
how the constraints on the potential parameters are altered when the initial
conditions are altered.

Throughout this paper we will assume a flat geometry for the Universe based on the location of the first peak in the
CMB angular power spectrum \cite{Netal02,Petal02} and work in reduced Planck units,
$M_{PL} = 1/\sqrt{8\pi G } = \hbar = c = 1$.
Whenever we require numerical values for the cosmological parameters we choose
$\Omega_{M}=0.3$ \cite{T02} and $H_{0}=72\;{\rm km/s/Mpc}$ \cite{Fetal01}.


\section{Tracker Quintessence} \label{sec:tracker}

Our focus in this paper is initial conditions and how they affect of Quintessence at late times.
Since we will frequently make comparisons of the solution calculated from arbitrary initial conditions
with the `tracker' solutions we begin with a derivation of the properties of the latter.

As with all the non-interacting fluid components in the Universe, the rate of
change of the quintessence energy density, $\rho_{Q}$, obeys
\begin{equation}
\dot{\rho}_{Q} = - 3\,H\,(\rho_{Q} + P_{Q}) \label{eq:rhodot}
\end{equation}
where $P_{Q}$ is the pressure of $Q$, the dot represents differentiation with
respect to cosmic time $t$, $H$ is the Hubble parameter,
\begin{equation}
H^{2} = \left( \frac{1}{a}\,\frac{da}{dt} \right)^{2} = \frac{\rho_{Q}+\rho_{f}}{3},
\end{equation}
$a$ is the scale factor and $\rho_{f}$ is energy density of all the other fluid components.
From this equation, and the definition of the energy density and pressure as
\begin{eqnarray}
\rho_{Q} & = & \frac{ \dot{Q}^{2} }{2} + V(Q) \label{eq:rhoQ}\\
P_{Q} & = & \frac{ \dot{Q}^{2} }{2} - V(Q), \label{eq:PQ}
\end{eqnarray}
the equation of motion for $Q$ is simply
\begin{equation}
\ddot{Q} + 3\,H\,\dot{Q} + V' = 0,  \label{eq:equation of motion}
\end{equation}
where $V'$ is the derivative of the potential with respect to $Q$.
From these general equations there are many different paths that would allow us to advance, each emphasizing
different properties of the field, the solution or the potentials (see, for example, \cite{SWZ99,ZWS99,LS99}).
With a general potential they are all equivalent of course. For us the direction we take is to
introduce the equation of state $w_{Q}=P_{Q}/\rho_{Q}$ as the ratio of the
quintessence pressure to its energy density and $\alpha=\rho_{Q}/\rho_{f}$ as the ratio
of the quintessence and remaining fluid energy densities. For these two quantities we can derive from equations
(\ref{eq:rhodot}) through (\ref{eq:equation of motion}) two first order, coupled, equations
\begin{eqnarray}
\frac{d\,\alpha}{da} & = & \frac{3\,\alpha\,(w_{f} - w_{Q})}{a}
\label{eq:dalphada} \\
\frac{d\,w_{Q}}{da} & = & \frac{3\,(1+w_{Q})\,(w_{Q} - v_{Q}^{2})}{a}
\label{eq:dwda}
\end{eqnarray}
where $w_{f}=P_{f}/\rho_{f}$ is the equation of state for the remaining fluid
components, $a$ is the scale factor, and $v_{Q}^{2}$ is simply
$v_{Q}^{2}=\dot{P}_{Q}/\dot{\rho}_{Q}$. The density parameter
$\Omega_{Q}$, the ratio of the quintessence total energy density to the critical density,
is related to $\alpha$ in our flat Universe by
\begin{equation}
\Omega_{Q}= \frac{\alpha}{1+\alpha}
\end{equation}
These equations are valid for any potential $V(Q)$.

For radiation and non-relativistic matter, $w$ and $v^{2}$ are constants, but
for the field this is not the case, and some remarkable properties of
quintessence can be attributed to this fact.
If we now adopt the inverse power law form for V
\begin{equation}
V = \frac{V_{0}}{Q^{\gamma}} \label{eq:IPL}
\end{equation}
then the speed $v_{Q}^{2}$ in equation (\ref{eq:dwda}) is the rather
complicated
\begin{equation}
v_{Q}^{2} = 1 - \frac{ \gamma\,(1-w_{Q}) }{ \sqrt{ 1+w_{Q}} }\;\sqrt{
\frac{\Omega_{Q}}{3} }\left[\frac{3\,\Omega_{Q}\,H^{2}\,(1-w_{Q})}{2\,V_{0}}
\right]^{1/\gamma}.
\label{eq:v2}
\end{equation}
Obtaining a solution for the evolution of the field is then a simple
case of integrating equations (\ref{eq:dalphada}) and (\ref{eq:dwda}),
with specified values for $\gamma$ and $V_{0}$ and initial values for
$\alpha$ and $w_{Q}$. These four parameters are not, however,
independent because the present state of the Universe is implicit in $H$ and
$w_{f}$.
To calculate $H$ we need to specify the energy densities of the fluid
components at all epochs: for the quintessence we can simply construct this
quantity according to equation (\ref{eq:rhoQ}), while for radiation and matter
this can be done by their simple scaling with $a$ along with knowledge of
their present energy densities. The contributions to the current radiation
density include the measured temperature of the CMB, and the calculated energy
density for three, non-degenerate neutrino flavors. This leaves only the present
matter density, which is usually expressed as the product of its density parameter
$\Omega_{M}$ and the present critical density $3\,H_{0}^{2}$. If we impose
the requirement of flatness then we must have $\Omega_{Q} = 1 - \Omega_{M}$
at the present time. We have thus specified the value that $\Omega_{Q}$ must
reach at the end of our integration thereby introducing a constraint and
eliminating one degree of freedom. If we regard the initial values for
$\alpha$ and $w_{Q}$ as inputs, and take $\gamma$ ($V_{0}$) as the
remaining free variable, then we must adjust $V_{0}$ ($\gamma$) to
produce a self-consistent cosmology.

The equations (\ref{eq:dalphada}) and (\ref{eq:dwda})
for $\alpha$ and $w_{Q}$ are clearly non-linear but certain
cases are open to analysis. In particular, we can look at the situations when either
derivative is zero and then examine whether these conditions are stationary. From equation
(\ref{eq:dalphada}) we find that trivially $d\alpha/da = 0$ when $\alpha =0$ and obviously
this is stationary since all higher derivatives of $\alpha$ with respect to the scale factor vanish.
There is also a second solution that occurs when $w_{f}=w_{Q}$ independent of
the value of $\alpha$. If $w_{f}$ is a constant then to be stationary we require $w_Q$
to be also independent of the scale factor but, as we will show below, this only occurs in the
limit $\gamma \rightarrow \infty$.

For equation (\ref{eq:dwda}), $dw_{Q}/da = 0$ occurs when $w_{Q} = \pm 1$; the
$w_{Q}=-1$ case is clear from equation (\ref{eq:dwda}), while
the $w_{Q}=+1$ solution arises because $v_{Q}^{2}=+1$ in such circumstances.
These extrema for the equation of state are only stationary for
certain potentials because they require either constant zero kinetic energy ($w_{Q}=-1$)
or potential energy ($w_{Q}=+1$). For the IPL they are impossible to achieve exactly but $w_{Q} = \pm 1$
can be approached in the limit when $\rho_{Q}$ is dominated by either $\dot{Q}^{2}/2$ or $V$.
There is also a third solution that occurs when $w_{Q} = v_{Q}^{2}$, and neither has
unity magnitude. The equality only occurs for specific values of $\alpha$ and $w_{Q}$ and to be stationary this
third solution requires a balance between the scale factor dependence of the $\sqrt\Omega_{Q}$ and
$[\Omega_{Q}\,H^{2}]^{1/\gamma}$ terms in equation (\ref{eq:v2}), i.e.
\begin{equation}
\sqrt{\Omega_{Q}}\;\left[ \Omega_{Q}\,H^{2} \right]^{1/\gamma} = constant.
\end{equation}
This result is equivalent to equation (5) of \cite{SWZ99}.
Inserting the definition for $\alpha$ this transforms into
\begin{equation}
\frac{\alpha^{1+2/\gamma}}{1+\alpha} =  \frac{\alpha_{I}^{1+2/\gamma}}{1+\alpha_{I}} \;\left( \frac{a}{a_{I}}
\right)^{6\,(1+w_{f})/\gamma}
\label{eq:constant}
\end{equation}
with the subscript $I$ denoting initial values.
But if $w_{Q}$ and $w_{f}$ are constant then from equation (\ref{eq:dalphada}) we also have
\begin{equation}
\alpha = \alpha_{I}\;\left(\frac{a}{a_{I}}\right)^{3\,(w_{f}-w_{Q})}.
\label{eq:tracker alpha}
\end{equation}
Thus we have two equations involving $\alpha$, (\ref{eq:constant})
and (\ref{eq:tracker alpha}), which must be self consistent in their scaling
with $a$. Taking the limit $\gamma \rightarrow \infty$ in equation
(\ref{eq:constant}) leads to $\alpha$ varying very slowly with the
scale factor, and therefore from equation (\ref{eq:tracker alpha}),
$w_{Q} \rightarrow w_{f}$. This result validates
our previous statement regarding the stability of $d\alpha/da=0$.
Alternatively, taking the limit $\alpha, \alpha_{I} \ll 1$ in equation
(\ref{eq:constant}) and equating it with equation
(\ref{eq:tracker alpha}) produces the requirement that
\begin{equation}
w_{Q} = \frac{\gamma\,w_{f} -2}{\gamma+2}. \label{eq:trackerw}
\end{equation}
for this case. As noted by Steinhardt, Wang and Zlatev \cite{SWZ99},
$w_{Q} \leq w_{f}$ so that from equation (\ref{eq:tracker alpha}) we see that $\alpha$ increases
with the evolution of the Universe. These solutions where $dw/da=0$ and
$\alpha \ll 1$ are the `tracker' solutions
originally named by Steinhardt, Wang and Zlatev \cite{SWZ99} and Zlatev, Wang and Steinhardt \cite{ZWS99}. As shown in
\cite{RP88,LS99}, these solutions are stable attractors for all positive values of
$\gamma$ in that perturbations away from this solution are damped. As we stated explicitly,
this epoch of constant $w_{Q}$ only occurs when $w_{f}$ is also constant so if we allow
more than one component for the background fluid then whenever the Universe
transits from domination by one component to another there will be a period
during which $dw_{Q}/da \neq 0$. Finally, if $w_{Q}$ is constant during an epoch,
the energy density of $Q$ scales as $\rho_{Q} \propto a^{-3\,(w_{Q}+1)}$ from which, after using the definition of
the equation of state, equation (\ref{eq:trackerw}) and the scale factor-time relation, we can obtain
\begin{equation}
Q = Q_{I} \; \left( \frac{t}{t_{I}} \right)^{2/(\gamma+2)}
\label{eq:Q(t)}
\end{equation}
where $Q_{I}$ is the value of the field at the initial time $t_{I}$.
This is the result obtained by Liddle and Scherrer \cite{LS99}.

For quintessence to be a viable explanation for the current epoch of
acceleration, it must have recently come to dominate the energy density of
the Universe, i.e. the Universe is transiting between matter domination to
dark energy domination. In such circumstances the analytic solution in
equation (\ref{eq:Q(t)}) is no longer valid and the equation of state is no
longer a constant. But there is still a solution to which all others converge
and we shall continue to refer to this solution as the `tracker' even if the
evolution of $Q$ is no longer given by equation (\ref{eq:Q(t)}) and the
name no longer justified in its original sense. The appeal of IPL models of
quintessence is not that the analytic equation for the tracker is always
valid or that $dw_{Q}/da$ is always zero but rather that a wide range of
initial conditions converge to a common solution thereby avoiding any fine
tuning of the initial values of $\alpha$ and $w_{Q}$. We shall discuss
further how the tracker is approached in the next section, for now we will
assume that the initial conditions are exactly those of the tracker solution.

The tracker solutions correspond to specific initial values of $\alpha$
and $w_{Q}$, the latter stems from equation (\ref{eq:trackerw}) while the
former can be derived from the requirement $w_{Q} = v_{Q}^{2}$ and equation
(\ref{eq:v2}). Since the initial conditions
are fixed the tracker solution is a function of only one parameter, $\gamma$.
We start integrating the equations at an initial scale factor $a_{I}$ of $a_{I}=10^{-30}$: this
initial value for the scale factor is, of course, arbitrary and we have checked
that our results are not sensitive to its value.

\begin{figure}[htb]
\centering \epsfxsize=3.3in \epsfbox{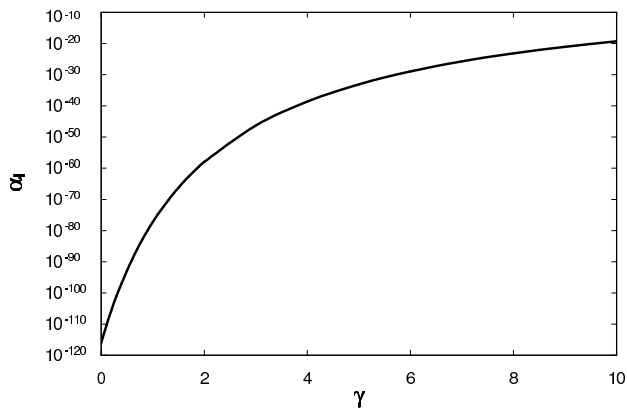}
\caption{The value of the ratio of the quintessence and background fluid
energy densities, $\alpha_{I}$, at the initial scale factor $a_{I}=10^{-30}$
as a function of the power law index $\gamma$.}
\label{fig:alphaI_versus_gamma}
\end{figure}
\begin{figure}[htb]
\centering \epsfxsize=3.3in \epsfbox{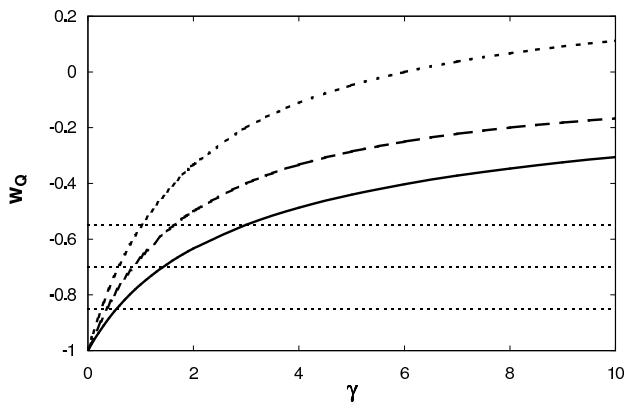}
\caption{The equation of state, $w_{Q}$, at the present time (solid), during matter domination
(long dashed) and during radiation domination (short dashed) as a function of the power law index $\gamma$.
The three horizontal, dashed lines correspond to the 1, 2 and 3-$\sigma$ limits of Bean \& Melchiorri \cite{BM01} }
\label{fig:final_w_tracker_versus_gamma}
\end{figure}

Figure (\ref{fig:alphaI_versus_gamma}) shows the initial tracker value of $\alpha$ at our choice
for the initial scale of $a_{I}=10^{-30}$ as a function of $\gamma$ The figure shows that $\alpha_{I}$
grows rapidly as $\gamma$ increases so that even modest values of
$\gamma$ raise the initial dark energy density by many orders of magnitude
above the cosmological constant value. In figure
(\ref{fig:final_w_tracker_versus_gamma}) we have plotted the equation of
state at the present time as a function of $\gamma$, together with the value
during both matter and radiation domination as given by equation
(\ref{eq:trackerw}). This figure can be used to derive an upper limit of $\gamma$
based simply on the requirement that the Universe be
accelerating at the present time. To accelerate in a flat Universe we must
satisfy $w_{Q}\,\Omega_{Q} \le -1/3$ so with
our adopted value of $\Omega_{Q}=0.7$ we find $\gamma \lesssim 5$.
If we use the Bean \& Melchiorri $3-\sigma$ limit of $w_{Q} \leq -0.55$ (though this was obtained by
assuming $w_{Q}$ is constant) then the limit drops to $\gamma \leq 3$.
The equation of state of the tracker at the present time is somewhat sensitive to the
adopted energy density $\Omega_{Q}$. Larger $\Omega_{Q}$ leads to slightly
smaller $w_{Q}$ for a given exponent so that enforcing $w_{Q}\,\Omega_{Q} \le -1/3$ leads to a
limit on $\gamma$ that rises slightly more quickly with $\Omega_{Q}$
than figure (\ref{fig:final_w_tracker_versus_gamma}) would seem to indicate.


\section{Non-Tracking Quintessence} \label{sec:nontracker}


\subsection*{Initial Conditions for the Q-field}

Most previous studies of cosmological constraints on the IPL model potential
have assumed that $Q$ reached the tracker long before the epoch at which the tests
are applied. This is justified since the appeal of the IPL model is the fact that
$Q$ reaches this solution for many initial conditions. Since the initial
conditions for $Q$ are the focus of this paper, it seems reasonable to ask what are
the `natural' initial conditions for $Q$.

Malquarti \& Liddle \cite{ML02} followed a quintessence field with an IPL
potential during inflation and determined that due to quantum fluctuations $Q$ will
diffuse to a very low energy density compared to the background fluid at the end
of inflation. In this scenario $Q$ is decoupled from all other fluid components,
including the inflaton. During inflation, the small scale set by $V_{0}$ in the $Q$
potential implies that the inflaton field dominated the dynamics of the universe.
Only for small values of $Q$ could the dynamics of the universe be affected by
quintessence. However, these authors found that at the end of inflation,
$\rho_{Q}$ is initially so small that the field does not reach the tracker until
the relatively late stages of cosmic evolution.

Alternatively we have initial conditions for $Q$ motivated from inflation where the
scalar inflaton field, $\phi$, transits through a period of kinetic energy
domination after inflation. The epoch in which the kinetic energy of the scalar
field dominates the expansion rate has been called `kination'\cite{J97}.
Previous authors have attempted to connect the scalar field in quintessence to the
field responsible for inflation, by transiting the field through a kinetic
energy dominated regime \cite{PV99,PR99,KR99,DV01}. If the scalar field, here
responsible for both inflation and quintessence (i.e. $\phi=Q$), does go through a
period of kination, this period must end by BBN so that the universe is radiation
dominated at this stage \cite{Ketal99}. This allows us to constrain the energy in $Q$ after
inflation. Starting from the scaling relation for temperature, $T \sim a^{-1}$, as well as
the scaling relations $\rho_{r} \sim a^{-4}$ and $\rho_{\phi} \sim a^{-6}$ during
kination, and expressing the relativistic energy as
$\rho_{r}={\pi}^2g_{*}T^{4}/30$, where $g_{*}$ counts the total number of
relativistic degrees of freedom, we obtain
\begin{equation}
\rho_{\phi,i}=\frac{\pi^{2}g_{*}}{30}\frac{T_{i}^{6}}{T_{*}^{2}}
\label{eq:phi_i}
\end{equation}
for the energy in $\phi$ at the end of inflation. Here $T_{i}$ and $T_{*}$ are the
temperatures at the end of inflation and at the end of kination, respectively,
and $g_{*} \sim 100$.  As noted in \cite{S93}, in order to satisfy the
nucleosynthesis bound, the temperature of particles created during kination just
after thermal equilibrium is achieved is $T_{eq} \gg 10^{6}$ GeV. Since
$T_{eq} < T_{i}$, we can take $10^{6}$ GeV to be a very generous lower bound for
$T_{i}$, which in (\ref{eq:phi_i}) gives $\rho_{\phi,i} > 10^{42}\;{\rm GeV^{4}}$. Thus
it seems prudent to consider initial conditions on $Q$ such that
$\rho_{Q} \sim \rho_{r}$ in the early universe.


\subsection*{Approaching the Tracker}

In has long been known that the tracker solution in IPL models (by which we mean the solution given by equation
(\ref{eq:Q(t)}) when $\Omega_{Q}$ is small or what this solution becomes when $\Omega_{Q}$ is large) is
an attractor but that under certain initial conditions the time required to reach the tracker solution
will be longer than the present age of the universe \cite{WF00,SWZ99}.
Several authors have derived limits on $\gamma$ based on the
argument that the tracker must be reached at some point before the present epoch.
For example: Steinhardt, Wang \& Zlatev \cite{SWZ99} bounded $\gamma \geq 5$
by demanding that the field reaches its tracker solution by matter-radiation equality when
initially the energy density of the Universe at the inflationary scale was equipartitioned between the Quintessence and
the remaining fluid components.
Similarly, Frieman and Waga \cite{FW98} obtained the bound $\gamma \geq 3.6$ using SN
Ia data for currently acceptable values of the matter density, $\Omega_{M} \sim 0.2-0.4$.
We shall not impose these restrictions and consider in this section all solutions that are able to satisfy
the boundary condition whether they reach the tracker by the present epoch or not. Since we shall
frequently compare a solution calculated from arbitrary initial conditions to the tracker
we will denote any quantity associated with the latter by the subscript $T$.

We introduce the quantity $u$, as in \cite{LS99}, defined to be the ratio of the
solution $Q(a)$ given arbitrary initial conditions to the tracker solution
$Q_{T}(a)$ for some fixed value of the IPL exponent $\gamma$ i.e.
\begin{equation}
u(a) = \frac{Q(a)}{Q_{T}(a)}.  \label{eq:u(a)}
\end{equation}
With this definition of $u(a)$ a solution reaches the tracker solution when $u=1,\;du/da=0$.
When $\Omega_{T} \ll 1$ we can derive an equation of motion \cite{LS99} from equation (\ref{eq:equation of motion})
after using equation (\ref{eq:Q(t)})
\begin{eqnarray}
t^{2}\,\ddot{u} & + & \left[\frac{4}{\gamma+2} + \frac{2}{w_{f}+1} \right]\,t\,\dot{u} \label{eq:u equation of motion} \\
& & -  \frac{2}{\gamma+2}\,\left[ \frac{\gamma}{\gamma+2} - \frac{2}{w_{f}+1} \right]\,
 \left( u - \frac{1}{u^{\gamma+1}} \right) =0. \nonumber
\end{eqnarray}

\begin{figure}[b]
\centering
\epsfxsize=3.3in \epsfbox{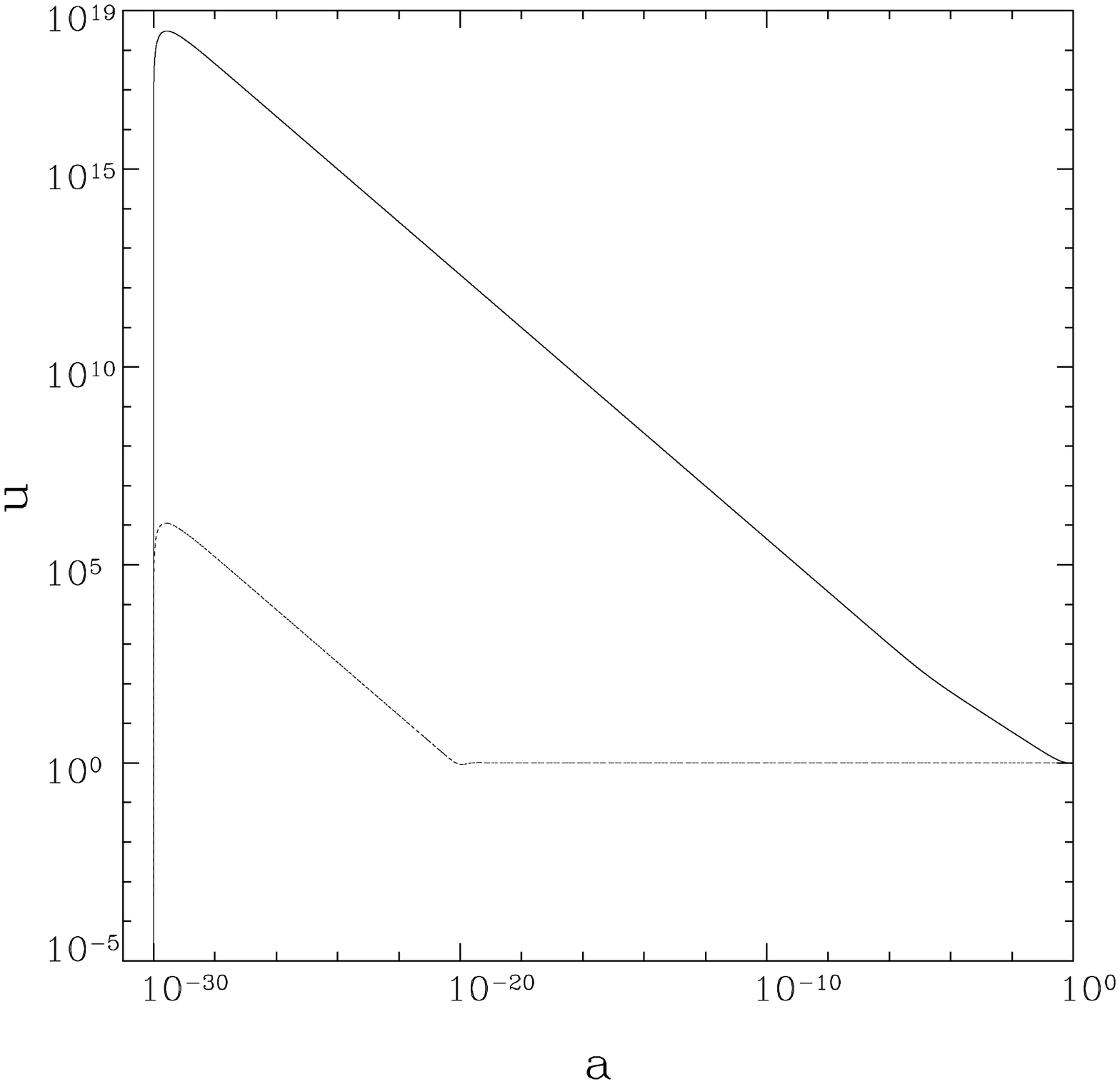}
\epsfxsize=3.3in \epsfbox{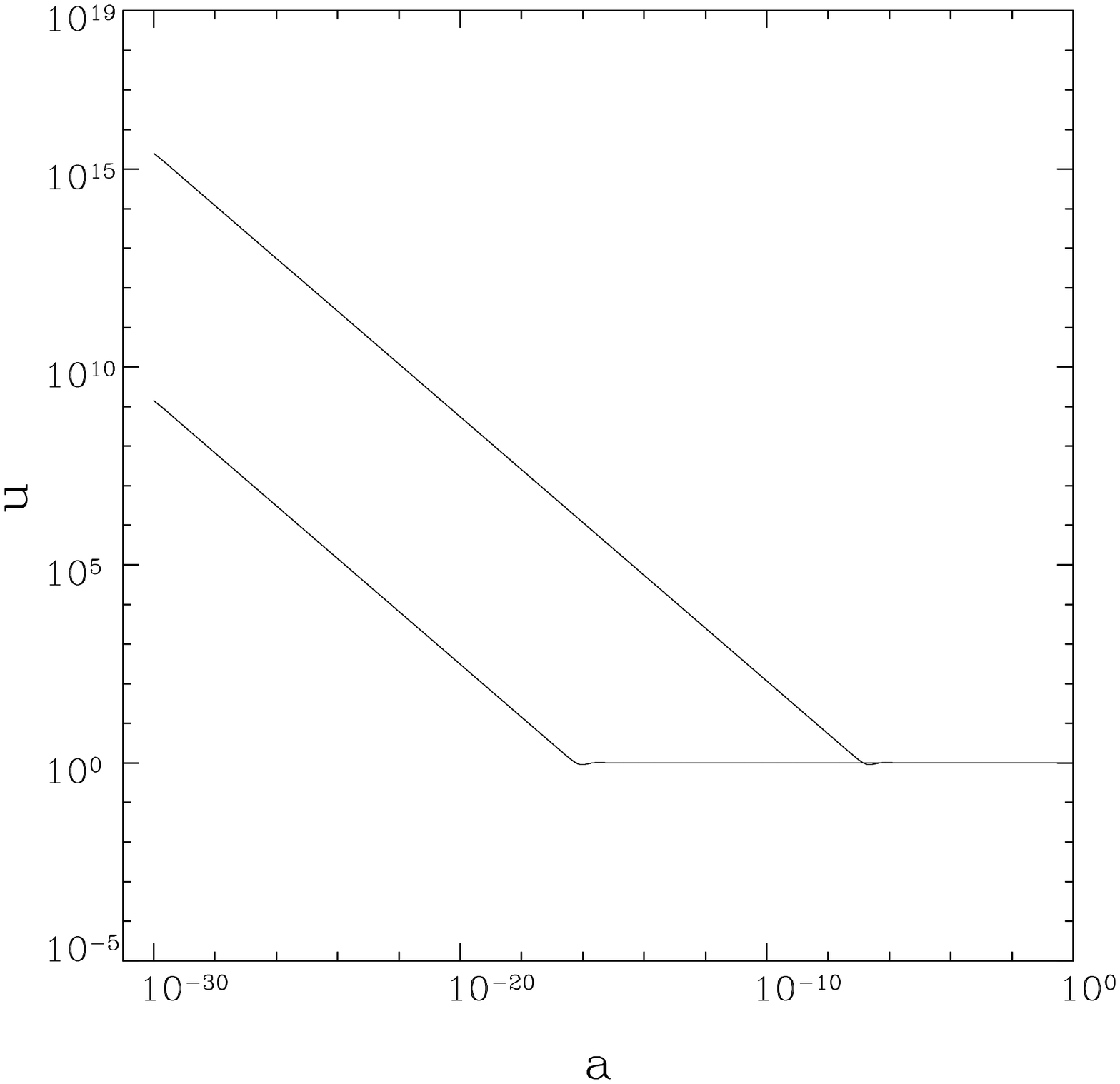}
\caption{The ratio of the value of
$Q$ to the tracker as a function of scale factor. The four curves
correspond to different initial values of $\alpha_{I}$, the ratio of the initial Quintessence energy density
to the background (radiation) component. In the top panel, from top to bottom, $\alpha_{I} \in \{ 1, 10^{-25}\}$,
while in the lower panel, from top to bottom, $\alpha_{I} \in \{10^{-100}, 10^{-75}\}$.}
\label{fig:u a}
\end{figure}
In figure (\ref{fig:u a}) we plot $u(a)$ for the case $\gamma=4$ for four values of
$\alpha_{I}$, the value of $\alpha$ at the beginning of the integration, from
$\alpha = 1, 10^{-25}, 10^{-75}, 10^{-100}$. In all cases we start the field from rest and at an
initial scale factor of $a_{I}=10^{-30}$.
From figure (\ref{fig:alphaI_versus_gamma}) we see that the tracker solution for
$\gamma=4$ corresponds to a value of $\alpha_{T} \sim 10^{-40}$ at $a_{I}=10^{-30}$ so our
demonstration values span a large range around the tracker.
Despite this large range for $\alpha_{I}$, both above and below
$\alpha_{T}$, the figure shows that the solutions for all are
remarkably similar in their general shape and in every case we have the
remarkable result that the tracker solution is always approached from
above. The oscillations around $Q_{T}$ seen in Liddle \&
Scherrer \cite{LS99} are too small to be seen with the scale of the
figure because they occur when $u \sim 1$.
For the two cases where $\alpha_{I} \leq \alpha_{T}$ the initial value of $Q$ is
much larger than the tracker $Q_{T}$ thus the initial $u$ is much greater than
unity. As the Universe evolves $u$ falls as a power law with $a$ and the two
solutions join the tracker at approximately a scale factor of $a \sim 10^{-16}$
for $\alpha_{I}=10^{-75}$, $a \sim 10^{-7}$ for $\alpha_{I}=10^{-100}$. In
contrast, when $\alpha_{I} \geq \alpha_{T}$ the initial value of $u$ is smaller
than unity. As we integrate the equations we see a very rapid
and immediate increase as $u$ rushes up to a peak value after which $u$ turns over
and again falls as a power law. For the two cases here the solution joins the
tracker at $a \sim 10^{-20}$ for $\alpha_{I}=10^{-25}$, $a \sim 1$ for
$\alpha_{I}=1$. From these four cases it is apparent that a value of
$\alpha_{I}$ in relative proximity to the tracker solution $\alpha_{T}$
joins the tracker very quickly but as $\alpha_{I}$ strays from
$\alpha_{T}$ the time required to reach the tracker grows rapidly. The
figure clearly shows that the epoch at which the tracker is joined can
easily be after the epoch of BBN, of recombination, or indeed it may
still not have happened.

\begin{figure}[t]
\centering
\epsfxsize=3.3in \epsfbox{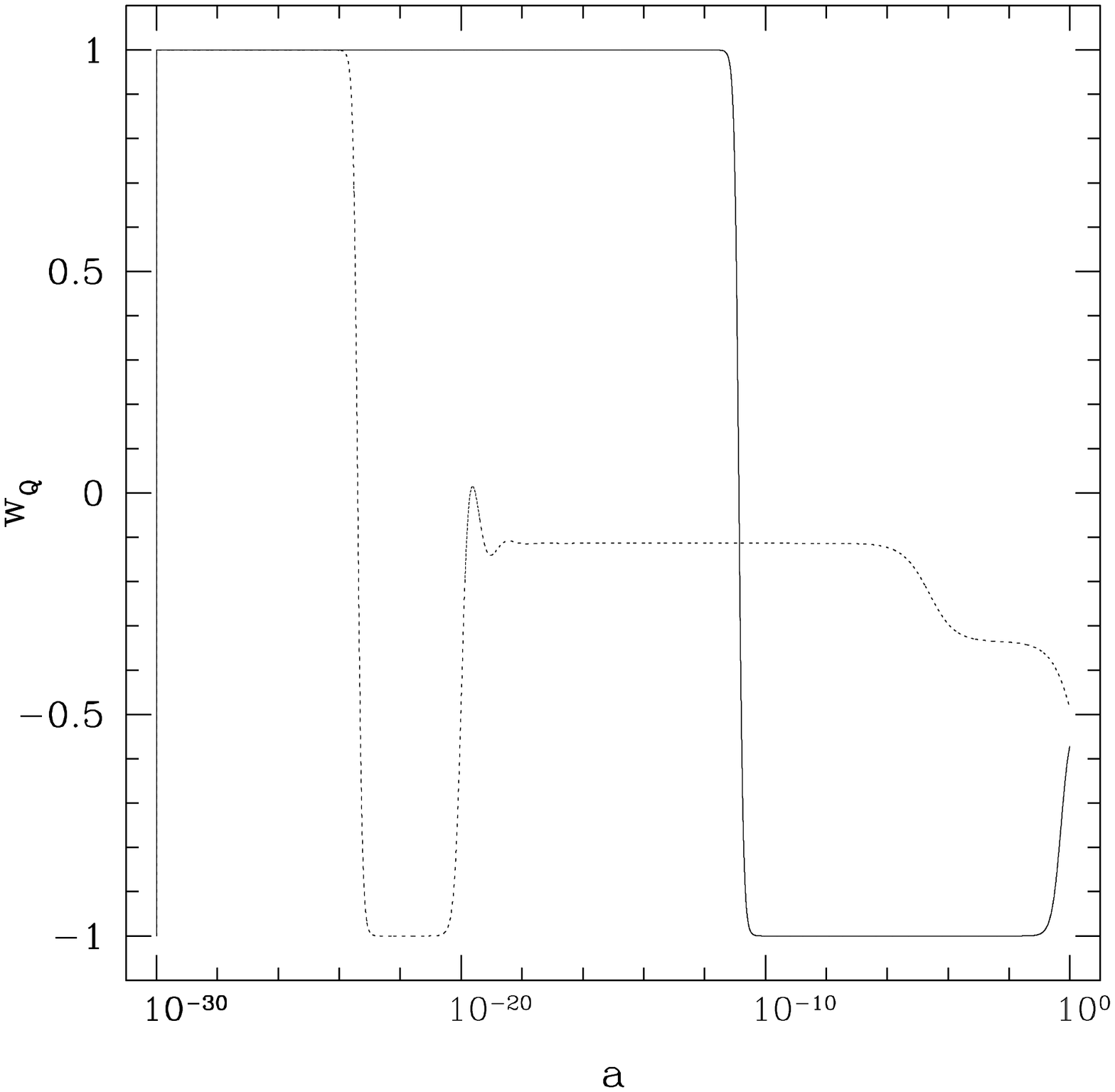}
\epsfxsize=3.3in \epsfbox{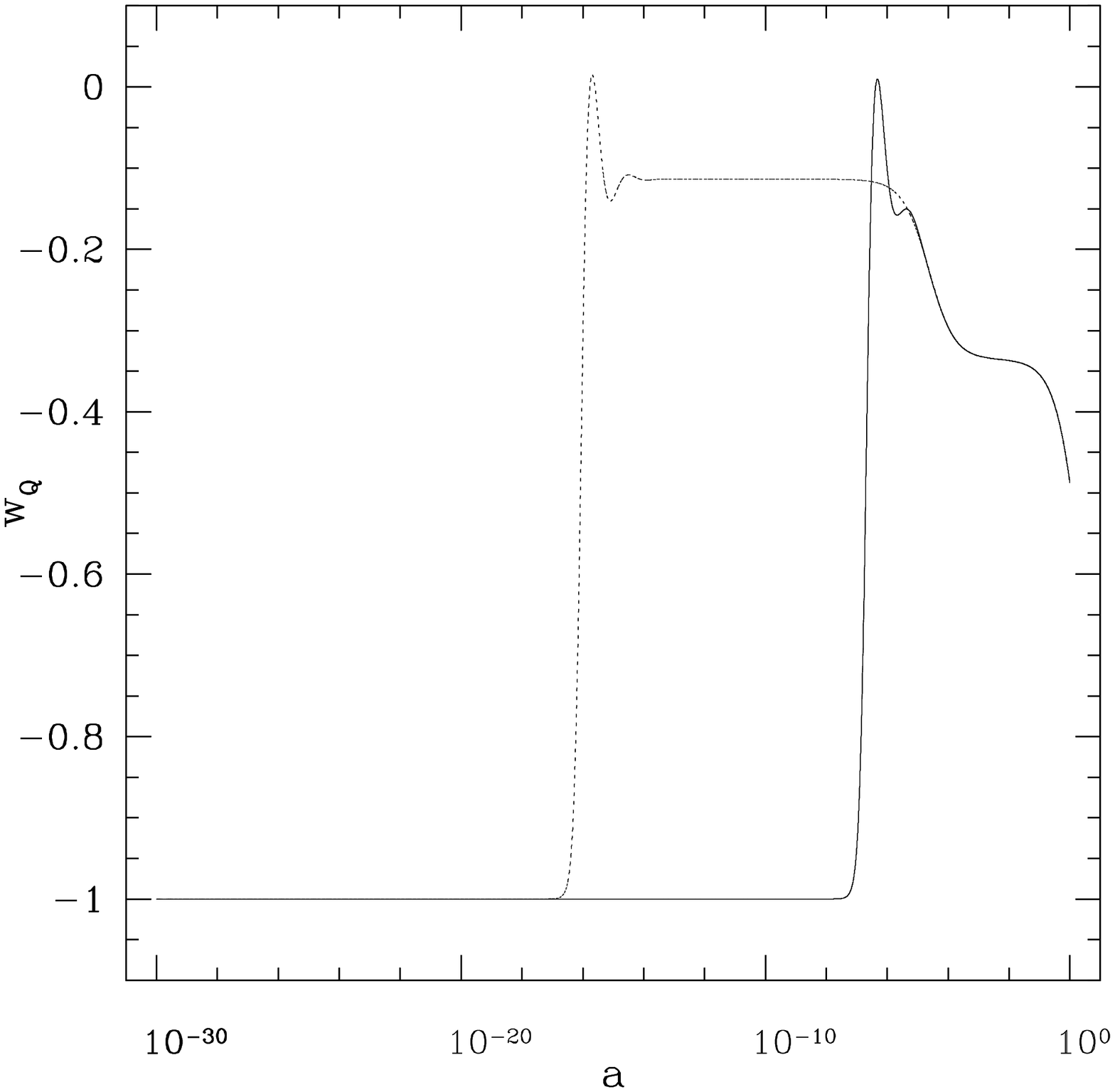}
\caption{The evolution of $w_{Q}$ as a function of scale factor for the values of $\alpha_{I}$ in figure
(\ref{fig:u a}). In the top panel, $\alpha_{I} = 1$ (solid) and $\alpha_{I}=10^{-25}$ (dotted) while in
the lower panel. $\alpha_{I} = 10^{-100}$ (dotted), $\alpha_{I}=10^{-75}$ (solid).
For $\alpha_{I} = 1$, the tracker is not reached by the present epoch and in this case $w_{Q}$
is only beginning its climb to the tracker presently.}
\label{fig:w a}
\end{figure}
In figure (\ref{fig:w a}) we plot $w_{Q}$ versus scale factor for $\gamma = 4$ and
the same four values of $\alpha_{I}$. The behavior of $w_{Q}$ in non-tracker phases for $Q$ within
the IPL model was also discussed by Zlatev, Wang and Steinhardt \cite{ZWS99}. This figure helps
explain the results in figure (\ref{fig:u a}). Every case is initially set at $w_{Q}=-1$ and for
$\alpha_{I} \leq \alpha_{T}$ the equation of state remains at this value until the
tracker is joined at $a \sim 10^{-16}$ and $a \sim 10^{-7}$ seen previously. Note that
in this figure the oscillations of the field around the tracker are seen as the small wiggles
in $w_{Q}$ immediately prior to the joining of the tracker solution.
During this phase of $w_{Q}=-1$ the field behaves as a cosmological constant. The
figure indicates that the larger the discrepancy between $\alpha_{I}$ and
$\alpha_{T}$ for $\alpha_{I} \leq \alpha_{T}$ the longer this pseudo-cosmological
constant phase and eventually we would expect that at some sufficiently small
$\alpha_{I}$ the solution never joins the tracker. However, such a small value of
$\alpha_{I}$ circumvents the main appeal of Quintessence and we will not pursue
this possibility further. In contrast, for those cases where $\alpha \geq \alpha_{T}$
the equation of state jumps from $w_{Q}=-1$ to
$w_{Q}=+1$ very rapidly and remains there for some time before promptly plummeting
to $w_{Q}=-1$. This period of kinetic energy domination is much longer than the
rise and turnover seen in figure (\ref{fig:u a}) and its duration
grows as the difference between $\alpha_{I}$ and $\alpha_{T}$ increases. Once
again, during the phase of $w_{Q}=-1$ the field behaves as a cosmological constant
and from there the solution joins the tracker in the same fashion as the two cases
where $\alpha \leq \alpha_{T}$. Eventually we expect that at some sufficiently
large $\alpha_{I}$ the equation of state never falls to $w_{Q}=-1$ and so the
limits on $w_{Q}$ exclude values of $\alpha_{I}$ larger than this. From our
numerical simulations we find that $w_{Q} \geq 0$ at the present time for
$\alpha_{I} \gtrsim 10^{83}$ when $\gamma=4$.

The solution for $u$ in the limit $u \gg 1$ is very different from that of Liddle \& Scherrer \cite{LS99}.
When $u \gg 1$ the $1/ u^{\gamma+1}$ term in equation (\ref{eq:u equation of motion})
is negligibly small and can be dropped. The solution of the equation is then the sum of two power laws
\begin{equation}
u = u_{+}\,\left( \frac{t}{t_{I}} \right)^{\lambda_{+}} + u_{-}\left( \frac{t}{t_{I}} \right)^{\lambda_{-}}
\label{eq:u(t)}
\end{equation}
where the exponents are
\begin{equation}
\lambda_{\pm} = \left( \frac{w_{f}-1}{2\,(w_{f}+1)} - \frac{2}{\gamma+2} \right)\,
\mp \,\left( \frac{w_{f}-1}{2\,(w_{f}+1)} \right).
\end{equation}
Since $w_{f} \leq 1$ the two exponents are both negative and $\lambda_{+} > \lambda_{-}$. In terms of the
initial conditions the two constants, $u_{+}$ and $u_{-}$, are given by
\begin{eqnarray}
u_{+} & = & u_{I} - u_{-} \label{eq:u plus} \\
u_{-} & = & \left( \frac{w_{f}+1}{w_{f}-1} \right) \, \left[ \dot{u_{I}}\,t_{I} + \frac{2 u_{I}}{\gamma+2} \right]
\nonumber \\
      &   & = \left( \frac{w_{f}+1}{w_{f}-1} \right) \,\frac{\dot{Q_{I}}\,t_{I}}{Q_{I}}\,u_{I} \label{eq:u minus}
\end{eqnarray}
From the second result in equation (\ref{eq:u minus}) we see $u_{-} < 0$. At late times only the first
term in (\ref{eq:u(t)}) is important and, after referring to equation (\ref{eq:Q(t)}), we see that its time
dependence is that of the tracker. The field $Q$ is fixed. If we start the field from rest then $u_{-} =0$
and so $u_{+} = u_{I}$. This is the behavior see in the bottom panel of figure (\ref{fig:u a}).

From examining the top panel of figure (\ref{fig:u a}) we see that empirically this same formula also
applies to the cases when $\alpha_{I} > \alpha_{T}$. Thus we are led to conclude that $Q$ is
essentially stationary even though the field may be dominated by its kinetic energy.
Such behavior is expected since nothing
in the derivation of equation (\ref{eq:u(t)}) depended upon $w_{Q}$, the only requirement was that $u \gg 1$.
However, in these cases the initial value of $u$ is much smaller
than unity so strictly equation (\ref{eq:u(t)}) does not apply exactly at $t_{I}$ and $u_{+}$ and $u_{-}$ are
therefore not given by equations (\ref{eq:u plus}) and (\ref{eq:u minus}).
But this limit is reached very quickly and, furthermore, the peak value of $u$ is approximately the same as
$u_{+}$, the normalization of the late time evolution. By locating the time $t_{\star}$ where the derivative of
$u$ vanishes we find that
\begin{equation}
u_{+} \approx u_{\star} = \left(\frac{\gamma+2}{2}\right) \,\frac{\dot{Q_{\star}}\,t_{\star}}{Q_{T\star}}.
\label{eq:u star}
\end{equation}
where the subscript $\star$ indicates all quantities are evaluated at $t_{\star}$.
First, we can obtain an estimate for $\dot{Q_{\star}}$ by assuming the Quintessence energy
density at the end of the initial ascent is approximately equal to the initial Quintessence energy, $\rho_{I}$, and,
furthermore, that it is dominated by its kinetic component. Under these approximations
$\dot{Q_{\star}} \approx \sqrt{2\,\rho_{I}}$.
Second, the peak is located so close to the beginning of the integration that $t_{\star} \approx t_{I}$
and $Q_{T\star} \approx Q_{TI}$ so if we make use of $t_{I} \sim 1/H_{I}$ and the definition of
$H$ the result is that
\begin{equation}
u_{+} \approx u_{\star} \approx \left(\frac{\gamma+2}{2}\right)\, \frac{\sqrt{\Omega_{I}}}{Q_{I}}\,u_{I} \equiv \frac{Q_{\star}}{Q_{I}}\,u_{I}
\end{equation}
where $\Omega_{I}$ is the initial Quintessence density parameter and we have introduced the quantity $Q_{\star}$.

\begin{figure}[th]
\centering \epsfxsize=3.3in \epsfbox{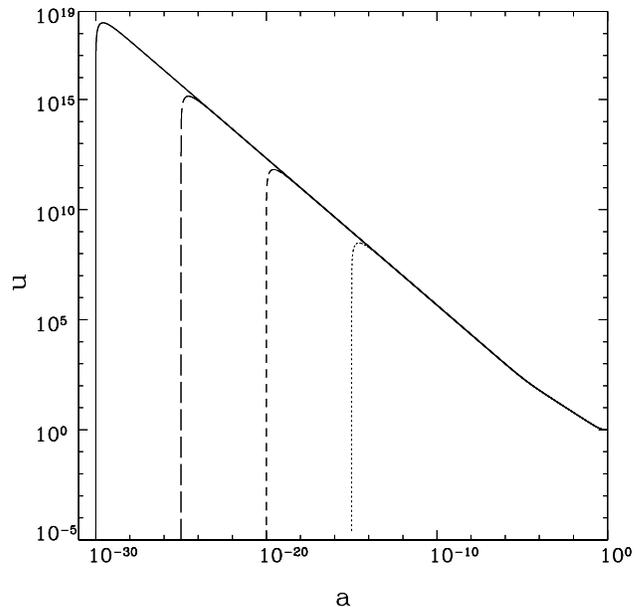}  \caption{The evolution of $u$
as a function of scale factor for $\gamma=4$, $\alpha_{I}=1$ and initial scale factors of
$a_{I} \in \{10^{-30},10^{-25},10^{-20},10^{-15}\}$ .}
\label{fig:u vs a, vary aI}
\end{figure}
Before proceeding to further investigate the implications of the non-tracker solutions we temporarily digress
to examine the importance of the initial scale factor upon our results.
So far we have begun all our integrations of $\alpha$ and $w_{Q}$ from an initial scale factor of $a_{I}=10^{-30}$.
We already know the tracker value of $\alpha$ increases with the scale factor from equation (\ref{eq:tracker alpha})
and as we showed in figures (\ref{fig:u a}) and (\ref{fig:w a})
the closer the value of $\alpha_{I}$ to $\alpha_{T}$ the shorter the non-tracker phase.
From these considerations it is therefore natural to expect solutions with an arbitrary value of
$\alpha_{I}$ to reach the tracker more rapidly as $a_{I}$ increases. Figure (\ref{fig:u vs a, vary aI}), a plot of $u$
versus $a$ when $\alpha_{I}=1$ and for different $a_{I}$, shows that this is exactly the case. As $a_{I}$ increases the
duration of the non-tracker phase and the height of the peak value of $u$ decrease.
But the figure also shows that the scale factor at which a solution
reaches the tracker is independent of $a_{I}$.
This is not totally unexpected: our estimate of
$Q_{\star} \approx \surd\,\Omega_{I}$ doesn't change with $a_{I}$ if we use the same value of
$\alpha_{I}$ so the change in $u_{\star}$ as $a_{I}$ increases simply reflects the evolution of the tracker solution.
From this result we see that the evolution of the field as we approach the current epoch is robust
and the necessity of specifying the initial scale factor for non-tracker solutions has disappeared.

When $\gamma=4$ the two cases $\alpha_{I} = 10^{-25}$ and $\alpha_{I} =10^{-75}$ are sufficiently
close to $\alpha_{T}$ that the tracker is reached during the radiation dominated
era, while the $\alpha_{I} = 10^{-100}$ curve reaches the tracker at the beginning
of the transition to the matter dominated era. The $\alpha_{I} = 1$ curve does not
reach the tracker at all, and were it not for the initial climb to $w_{Q} = +1$,
this case would be identical to a cosmological constant. This is consistent with our
expectations if $Q_{\star}$ were approximately the same as the present value of $Q_{T}$.
From Malquarti \& Liddle \cite{ML02} $V_{0} \sim 10^{-120}$ (in reduced Plank mass units)
for $\gamma=4$ so together with $w_{T} \sim -1$ for the current equation of state of the tracker
and our concordance values of $\Omega_{Q}$ and $H_{0}$
then $Q_{T} \sim \;{\cal O}(1)$. Thus, $Q_{\star}$ must also be of this order implying
that $\alpha_{I}  \gtrsim 1$ in order for the solution to avoid joining the tracker by the present epoch.
\begin{figure}[t]
\centering \epsfxsize=3.3in \epsfbox{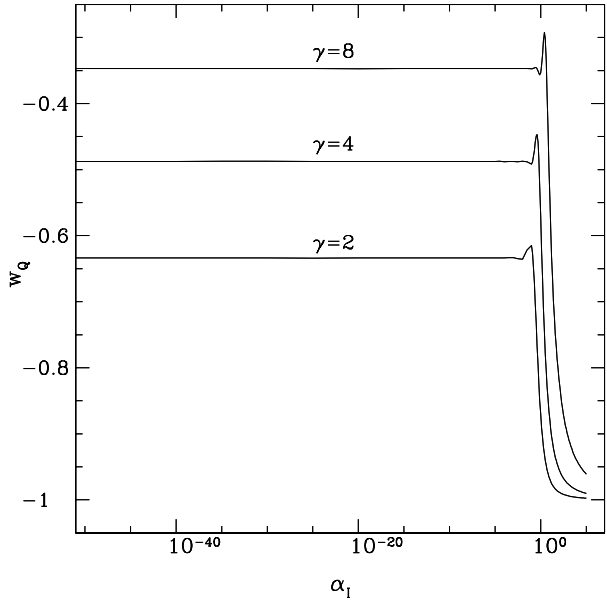}
\caption{The final value of the equation of state as a function of $\alpha_{I}$.
The slight jump at $\alpha_{I}=1$ is a result of the field oscillating about the tracker
today.}
\label{fig:final_w_vs_alpha}
\end{figure}
Identifying $\alpha_{I} \sim 1$ as dividing the cases where the tracker was
joined from those where it has not is confirmed (and is much more obvious) in
figure (\ref{fig:final_w_vs_alpha}) where we plot the
equation of state at the current epoch as a function
of $\alpha_{I}$ for different values of $\gamma$.
This figure clearly shows the change at $\alpha_{I} \sim 1$ where the
equation of state of $Q$ drops to $w_{Q} = -1$ as $\alpha_{I}$ passes through the
equipartition value. There is a slight movement of the transition point with $\gamma$ but it is not large.

\begin{figure}[th]
\centering \epsfxsize=3.3in \epsfbox{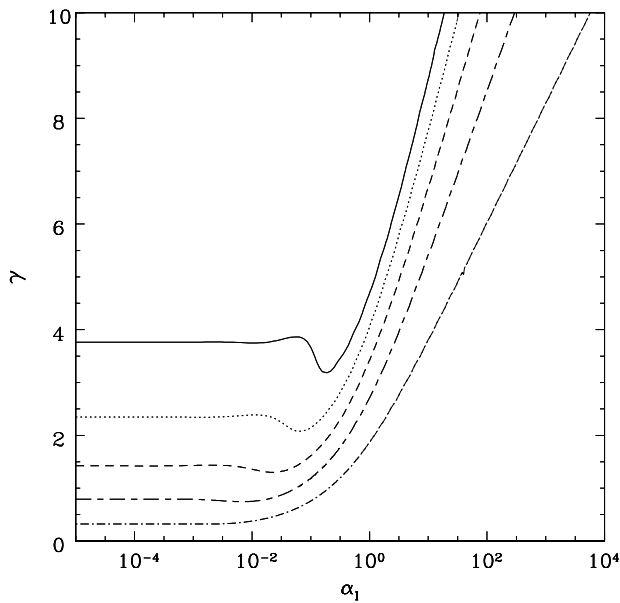}  \caption{Curves of
constant $w_{Q}$ at the present epoch as a function of $\gamma$ and $\alpha_{I}$. From top to
bottom, the curves represent $w_{Q} = -0.5,-0.6,-0.7,-0.8,-0.9$.} \label{fig:wcontourplot}
\end{figure}
In figure (\ref{fig:wcontourplot}) we summarize our discussion
of non-tracker Quintessence with a contour plot of the final value of
$w_{Q}$ as a function of $\gamma$ and $\alpha_{I}$. For values of
$\alpha_{I} \lesssim 1$, the final value of $w_{Q}$ is only a function
of $\gamma$ while for $\alpha_{I} \gtrsim 1$ we have the situation
where the equation of state at the present time is close to $-1$ for a
large range in $\gamma$. The slight dip in the contours of $w_{Q} =
-0.5,-0.6,-0.7$ are a result of $Q$ just reaching the tracker at the
present epoch. Just before $Q$ `hooks' the tracker it oscillates around
that solution \cite{LS99} and these oscillations are seen as dips in
the contours. The figure shows
essentially a dichotomy of acceptable solutions in the IPL model. In
order to obtain an equation of state within the current observationally favored
limits $-0.85 \geq w_{Q} \geq -1$ we are forced to small values of $\gamma$ if
$\alpha_{I} \lesssim 1$ \emph{but} if the initial conditions are such that the
quintessence energy density is comparable to, or larger than, the radiation
component virtually any value of $\gamma$ is permitted. A Universe initially
dominated by a quintessence field evolves to one that effectively possesses a
cosmological constant, independent of the value of $\gamma$.


\section{Quintessence, The CMB, and SN Ia} \label{sec:CMBSN Ia}

We have already derived an upper limit to $\gamma$ just from the requirement that the
Universe accelerate. We can improve upon this result by using observations
of the CMB and SN Ia, which provide powerful probes of cosmology from
a redshift of $\sim 1000$ and lower.

The significant changes introduced into the CMB angular power spectrum by
replacing a vacuum energy with an IPL Quintessence field have been discussed by
many authors including Ratra \& Peebles \cite{RP88} and Kneller \& Steigman
\cite{KS03}.
The position of the peaks in the
angular power spectrum are proportional to the ratio of the sound horizon at
recombination to the angular diameter distance at that redshift, while the height
of the peaks is set by their variance relative to the variance at the largest
angular scales as measured by COBE.
For a given redshift and matter density, Quintessence increases the dark
energy density, leading to a swifter expansion rate. For the IPL tracker solution, the
contribution from $\rho_{Q}$ is small until the recent epoch, thus its
influence only becomes significant at low redshifts. The swifter expansion results in a
reduction of the angular diameter distance,
\begin{equation}
d_{A}(z)=(1+z)^{-1}\int_0^{z},dz'/H(z')
\end{equation}
while leaving the sound horizon at recombination unaltered. An increase in $H(z)$ therefore moves
the peak positions to larger angular scales and they become more tightly
bunched. This effect increases with $\gamma$ because for tracker solutions
the energy density is correlated with the
exponent: smaller $\gamma$ corresponding to smaller $\rho_{Q}$ at any fixed redshift.
At the same time models with Quintessence always pass through the matter-dark
energy transition at a higher redshift even if they typically begin to
accelerate at a lower redshift. The earlier dominance of the dark energy will
result in a more significant decay of the gravitational potentials so that any
photon traversing a potential well will emerge with a larger blueshift relative
to the background expansion. The increased blueshifting (and the increased
redshifting from potential hills) leads to an enhancement in the variance of the
CMB temperature. This integrated Sachs Wolfe (ISW) effect is most effective at
scales above the causal horizon at the time of the matter-dark energy transition
i.e. the largest angular scales. But the COBE normalization of the CMB power
spectrum is exactly at these scales so the net result is that it is the variance
at the smallest scales that is reduced. Once again, this effect increases with
$\gamma$ for exactly the same reason.

The recent evidence for the current acceleration of the universe comes
from the use of SN Ia as `standard candles' at high-redshifts, $z \sim 0.3-0.8$
\cite{SAG99}. SN Ia are good standard candles because of the strong correlation
between the peak in the absolute magnitude and the decline in brightness 15 days
after the peak \cite{P94,H95,RPK96}. A measurement of the apparent magnitude is all that is
needed to determine the luminosity distance $d_{l}$. This same quantity can be
computed for a given cosmology through
\begin{equation}
d_{l}=(1+z)\,\int_0^{z} \, \frac{dz'}{H(z')}.
\end{equation}
From this equation it is immediately obvious that an increase in the Hubble
parameter results in a smaller luminosity distance and hence a brighter SN Ia for
a given redshift. For tracker solutions, as $\gamma$ increases the luminosity distance decreases and so
the same SN Ia will appear brighter still.

\begin{figure}[tb]
\centering \epsfxsize=3.3in \epsfbox{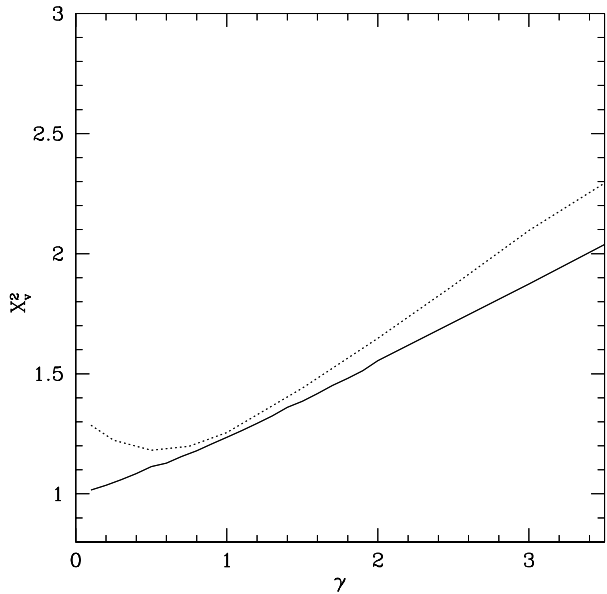}
\caption{The $\chi^{2}$ per degree of freedom for the SN Ia (dotted) and CMB
(solid) data, with $Q$ in the tracker solution. The `goodness-of-fit' limit for the
SN Ia data at 99\% is $\chi^{2}_{\nu}=1.50$ while for the CMB data the 99\% limit
is at $\chi^{2}_{\nu}=1.60$}
\label{fig:chis_tracker}
\end{figure}
In figure (\ref{fig:chis_tracker}) we compare the IPL model predictions using tracker initial conditions
with the CMB and SN Ia data by plotting $\chi^{2}_{\nu}$, the $\chi^{2}$ per degree of
freedom $\nu$ as a function of $\gamma$. For the
CMB data we use the 41 points of BOOMERANG \cite{BOOMERANG}, DASI \cite{DASI} and
MAXIMA \cite{MAXIMA} together with RADPACK \cite{RADPACK} to calculate the
$\chi^{2}$. We optimize over the scalar tilt and the baryon-to-photon ratio but,
in order to make the calculation computationally feasible, we have not
marginalized over the present Hubble parameter or the matter density $\Omega_{M}$.
For SN Ia, we use the 54 data points from
`fit C' of \cite{SAG99}. For the CMB data the 99\% goodness-of-fit limit is $\chi_{\nu}^{2} = 1.60$
while the equivalent limit when using the SN Ia data is $\chi_{\nu}^{2} = 1.50$. The
figure clearly shows that smaller values of $\gamma$ are favored and we can
extract an upper limit to $\gamma$ of $\gamma \leq 1.6$ at 99\% from the SN Ia data, $\gamma \leq 2.2$
at 99\% from the CMB data.
This limit is a somewhat stringent because we have not marginalized over all
the remaining parameters but the result is not very much different from other
studies of IPL Quintessence such as Malquarti \& Liddle \cite{ML02} who derived
$\gamma \leq 2$ at 99\%.
Figure (\ref{fig:chis_tracker}) clearly demonstrates that if we use tracker initial conditions then the field must be
very similar to the cosmological constant. Although we make this statement after
considering only the IPL model this result is in qualitative agreement with other
analysis of quintessence using the CMB and SN Ia which used a constant equation of
state $w_{Q}$ \cite{BHM01,CC02,HM02,E99,B01}.


\subsection*{CMB, SN Ia without the tracker}

As we have seen from figure (\ref{fig:wcontourplot}), $\alpha_{I}>1$ widens the
allowed parameter space in $\gamma$ when fitting $w_{Q}$ to present observed
bounds. Larger values of $\alpha_{I}$ behave more like a cosmological constant,
$w_{Q}=-1$, except for the caveat of an initial period when $w_{Q}=+1$. Thus,
since larger values of $\alpha_{I}$ push the current value of $w_{Q}\rightarrow-1$, we might expect
that these solutions would provide a very good fit to the CMB and SN Ia data.

\begin{figure}[tb]
\centering \epsfxsize=3.3in \epsfbox{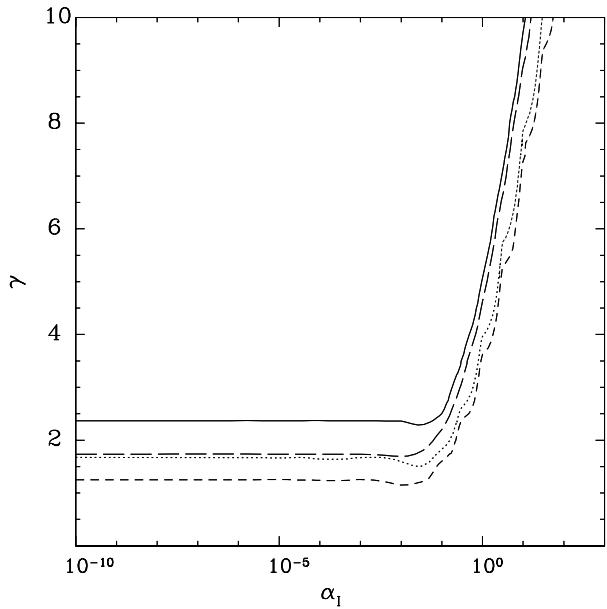}  \caption{Contours of
constant $\chi^2_{\nu}$ in the $\gamma-\alpha$ plane. From top to bottom, the
curves are: $\chi^{2}_{\nu}=1.60$ (solid) and $1.40$ (long dashed) corresponding to a
probability of $99\%$ and $95\%$ respectively for the fit between theory and the CMB data, $\chi^{2}_{\nu}=1.50$ (dotted) and $1.34$
(dashed) which are the 95\% and 99\% goodness-of-fit of the models to the SN Ia data.}
\label{fig:SNandCMBcontours}
\end{figure}
In figure (\ref{fig:SNandCMBcontours}) we plot the contours of constant
$\chi_{\nu}^{2}=1.34$ and $1.50$, that correspond to 95\% and 99\% goodness-of-fit to the SN Ia,
and $\chi_{\nu}^{2}=1.40$ and $1.60$, the 95\% and 99\% goodness-of-fit to the CMB data, again in the $\gamma-\alpha$ plane.
For $\alpha_{I} \lesssim 1$ the exponent $\gamma$ is very tightly bound by the CMB
and SN Ia because in this region the field joined the tracker solution before the
present epoch. The limits on $\gamma$ are exactly the same as found for tracker
initial conditions derived from figure (\ref{fig:chis_tracker}).
But when $\alpha_{I} \gtrsim 1$ the field has not joined the
tracker and we find the range of acceptable values of $\gamma$ increases
significantly such that much of the region $\alpha_{I}\gtrsim 1$ is compatible
with SN Ia and CMB constraints.
Thus the frequently stated limits on the exponent of power-law potential only
apply when the initial value of $\alpha_{I}$ falls within the range that will
converge to the tracker solution by the present day for a given value of $\gamma$.


\section{Conclusions}

We have examined the IPL quintessence model focusing on the influence of the
initial conditions upon the behavior of the solutions and in particular those
cases where the field has not converged to the tracker solution by the present
day. We have re-analyzed the previously derived bounds on the exponent $\gamma$
in the potential and find initial conditions that lead to solutions where the
field reached the tracker before the present day limits the exponent to $\gamma \lesssim 2$.
However, after we looked at initial conditions where the field does not reach the tracker we found much larger values
of $\gamma$ are permissible. The dividing line between the two types of solution
occurs at the point of equipartition between the initial Quintessence and
radiation energy densities and this is independent of the initial scale factor.
IPL models of Quintessence give rise to values of the equations of state at the
present time that are close to $-1$ either by the exponent being small, $\gamma \lesssim 2$, with small
initial energy densities relative to the radiation \emph{or} very large initial energy densities
relative to the radiation content and arbitrary exponents.


\acknowledgments

We wish to thank Bob Scherrer, Gary Steigman, Terry Walker, and Andrew Zentner for
useful discussions. We are grateful to Uros Seljak and Matias
Zaldarriaga for the use of the publicly available code CMBfast. This work was supported
in part by the DOE (DE-FG02-91ER40690), and by the Ohio State University
Department of Physics .


\end{document}